\documentclass[runningheads]{llncs}
\usepackage{longtable}
\usepackage{graphicx}
\usepackage{amsmath}
\usepackage{array}
\usepackage{xcolor}
\usepackage{multirow}
\usepackage{float}
\let\paragraph\oldparagraph
\let\subparagraph\oldsubparagraph
\usepackage{amsmath,amssymb,amsfonts}
\usepackage{algorithmic}
\usepackage{textcomp}
\usepackage{balance}
\usepackage[T1]{fontenc}

\DeclareMathOperator*{\argmin}{argmin}
\usepackage[ruled, noend, linesnumbered]{algorithm2e}
\usepackage{graphicx}
\usepackage[pagewise]{lineno}
\SetKwInput{KwInput}{Input}                % Set the Input
\SetKwInput{KwOutput}{Output} 
\usepackage{svg}
\usepackage{tabularx}
\usepackage{textcomp}
\usepackage{xcolor}
\usepackage{soul}
\usepackage{comment}
\usepackage{multirow}
\usepackage{romannum}
\usepackage{appendix}
\usepackage{subcaption}
\usepackage{titlesec}

\makeatletter
\newcommand{\printfnsymbol}[1]{%
  \textsuperscript{\@fnsymbol{#1}}%
}
\makeatother
\begin{document}
\title{Validating Optimal COVID-$19$ Vaccine Distribution Models}
%: A Constraint Programming Approach}
\author{Mahzabeen Emu\orcidID{0000-0002-0433-1873} \thanks{Equal contribution}\and
Dhivya Chandrasekaran\orcidID{0000-0001-7268-709X} \printfnsymbol{1}\thanks{Corresponding Author}\and
Vijay Mago \orcidID{0000-0002-9741-3463} \and
Salimur Choudhury \orcidID{0000-0002-3187-112X} }
\authorrunning{M. Emu et al.}
\institute{Department of Computer Science, Lakehead University, Thunder Bay, Ontario, Canada, P7B 5E1 
\email{\{memu,dchandra,vmago,schoudh1\}@lakeheadu.ca}}
\maketitle              % typeset the header of the contribution
\begin{abstract}
With the approval of vaccines for the coronavirus disease by many countries worldwide, most developed nations have begun, and developing nations are gearing up for the vaccination process. This has created an urgent need to provide a solution to optimally distribute the available vaccines once they are received by the authorities. In this paper, we propose a clustering-based solution to select optimal distribution centers and a Constraint Satisfaction Problem framework to optimally distribute the vaccines taking into consideration two factors namely priority and distance. We demonstrate the efficiency of the proposed models using real-world data obtained from the district of Chennai, India. The model provides the decision making authorities with optimal distribution centers across the district and the optimal allocation of individuals across these distribution centers with the flexibility to accommodate a wide range of demographics. 

\end{abstract}
\section{Introduction}
The ongoing pandemic caused by the Severe Acute Respiratory Syndrome Coronavirus 2 (SARS-CoV-2) called the coronavirus disease (COVID-19), has not only caused a public health crisis but also has significant social, political, and economic implications throughout the world \cite{ali2020covid,nicola2020socio,xiong2020impact}. More than 92.1 million cases and 1.9 million deaths have been reported worldwide as of January 2021 \cite{dong2020interactive}. One of the widely used solutions, in preventing the spread of infectious diseases is vaccination. Vaccination is defined by WHO as ``{\it A simple, safe, and effective way of protecting people against harmful diseases before they come into contact with them. It uses the body’s natural defenses to build resistance to specific infections and makes the immune system stronger” \footnote{https://www.who.int/news-room/q-a-detail/vaccines-and-immunization-what-is-vaccination}}. Various researchers and pharmaceutical corporations began research work on identifying potential vaccines to combat the spread of COVID-19. Of the number of vaccines under trial, Tozinameran (BNT162b2) by Pfizer and BioNTech, and mRNA-1273 by Moderna have achieved an efficacy of more than 90\% \cite{mahase2020covid,polack2020safety}. Various nations have begun the process of approval of these vaccines for mass distribution and have placed orders to facilitate a continuous flow of supply of vaccines to meet the demands.

In this paper we propose a Constraint Satisfaction Programming (CSP) framework based model to optimize the distribution of vaccine in a given geographical region. We aim to maximize the distribution of vaccine among the group of the population with higher priority while minimizing the average distance travelled by any individual to obtain the vaccine. In order to justify the efficiency of the proposed model we compare the performance of the following four optimization models namely,
\begin{itemize}
    \item Basic Vaccine Distribution Model (B-VDM)
    \item Priority-based Vaccine Distribution Model (P-VDM)
    \item Distance-based Vaccine Distribution Model (D-VDM)
    \item Priority in conjunction with Distance-based Vaccine Distribution Model (PD-VDM). 
\end{itemize}  
and present how the model PD-VDM provides the most optimal solution for the distribution of vaccines. We perform a Case Study using the demographic data obtained from Chennai - a well-renowned city in Southern India. This case study highlights how the model can be used by decision making authorities of a city with an population of 5.7 million. The model aids the decision making authorities to choose an optimal number of vaccine distribution centers (DCs), and to optimally assign an individual to a hospital such that the individuals in the priority groups are vaccinated first while minimizing the distance they travel to the vaccine DCs. In section \ref{relatedwork} we discuss the various steps and challenges involved in the process of vaccine distribution, in section \ref{methodology}  we describe in detail the procedure followed to build the proposed models. The case-study is discussed in section \ref{casestudy} followed by the discussion of results in section \ref{results}. We conclude with future research directions in section \ref{conclusion}.
\vspace{-3mm}
\section{Related Work} \label{relatedwork}
\vspace{-3mm}
In order to effectively distribute vaccines, it is necessary to understand the supply chain of vaccines. The supply chain of vaccines is divided into four major components namely the product, production, allocation, and finally the distribution \cite{duijzer2018literature}. The first concern of the decision making authorities is to decide on which vaccine to choose for distribution in their country, province, or region. For example, for countries in tropical regions, the storage temperature of the vaccine is an important factor. Similarly, while developed countries are able to afford vaccines at a higher price, most developing countries prefer the vaccine which has an affordable price. In the present scenario, three of the prominent vaccines in play are the BNT162b2,  mRNA-1273, and AZD1222 \cite{kaur2020covid}. They have storage temperatures of -70℃, -20℃, and 0℃ and cost USD 20, USD 50, and USD 4 respectively \footnote{https://www.nytimes.com/interactive/2020/science/coronavirus-vaccine-tracker.html}. Based on the storage it is safe to assume that while countries in temperate regions like the United States, Canada, and Russia would have the option of purchasing any one of these vaccines while tropical countries like India, Bangladesh, Pakistan would prefer AZD1222. The cost of the said vaccines also has a significant impact on the decision-making process of developing nations which mostly cater to a greater number of people. %GAVI, the vaccine alliance, established by the World Health Organization (WHO), caters to the immunization of various developing countries by forming alliances with them. GAVI, WHO, and Coalition for Epidemic Preparedness Innovations (CEPI) coordinate the COVAX program which is a “global collaboration to accelerate the development, production, and equitable access to COVID-19 tests, treatments, and vaccines.” %
Once the product is chosen, the production of vaccines has to be scheduled according to the demand. Factors that are taken into consideration at this stage include the production time, capacity for manufactures, supply-demand analysis, and so on \cite{begen2016supply}. Depending on the stage of the epidemic and the severity, the demand for the vaccine and the optimal timeline for the supply of vaccines may vary. Allocation and distribution stages go hand in hand in the vaccine supply chain. Depending on the distribution strategy, the allocation of vaccines at any level is tuned to achieve the best result. Allocation at a global level may depend on priority established through contracts and agreements among pharmaceuticals and governments. However, once a country receives the vaccines, further allocating the vaccines to provinces, states, or subgroups of the population is a critical decision that in turn has an impact on the distribution strategy. The distribution stage of the supply chain addresses the challenges of establishing an effective routing procedure, infrastructure of the vaccine distribution centers, inventory control, workforce, etc.

Operation Research (OR) involves the development and use of various statistical and mathematical problem-solving strategies to aid in the process of decision making. Various OR models are proposed over time to optimize the distribution of vaccines from the distribution centers. Ramirez-Nafarrate et al. \cite{ramirez2015point} proposes a genetic algorithm to optimize the distribution of vaccines by minimizing the travel and waiting time. Huang et al. \cite{Huang2017} formulated a vaccine deployment model for the influenza virus that ensures geographical priority based equity in Texas. However, their mathematical model might have generalization issues when applied to smaller or larger than state-level regions. Lee et al. \cite{lee2013advancing} developed the RealOpt\textsuperscript{\copyright} a tool to aid in identifying optimal location for vaccine distribution centers, resource allocation and so on. Researchers have also provided models to accommodate specific locations, for example Aaby et al. \cite{aaby2006montgomery} proposes a simulation model to optimize the allocation of vaccine distribution centers at Montgomery county, Maryland. This model aims to minimize the vaccination time and increase the number of vaccinations. While the above models consider the distribution centers to be stationary, Halper et al. \cite{halper2011mobile} and Rachaniotis et al. \cite{rachaniotis2012deterministic} consider the vaccine distribution centers to be mobile and address this as a routing problem. While the later model proposes that various mobile units serve different nodes in a network, the latter considers that a single mobile units serves various areas with a goal to minimize the spread of infection. Some of the OR models used in epidemics' control include non-linear optimization, Quadratic Programming (QP), Integer Linear Programming (ILP) and Mixed Integer Linear Programming (MILP) \cite{brandeau2005allocating}. 
ILP, MILP, and QP models are not suitable for many practical use cases due to its time expensive nature and infeasibility issues prevailing with irrational model designs. The mathematical design process and selection of pre-defined numerical bounds can lead to several technical glitches in the models. Despite the apparent similarities with ILP and MILP, CSP can eliminate all the aforementioned drawbacks and ensure sub-optimal solution in non-deterministic polynomial time by applying boolean satisfiability problem formulation. In the field of computer science, CSP is considered as a powerful mechanism to address research challenges regarding scheduling, logistics, resource allocation, and temporal reasoning \cite{CSP-Survery}. Hence in this article we employ CSP to propose four models to optimize the distribution minimizing the traveling distance and maximizing the vaccination of high priority population.
%To be included in discussion section: Many researchers may prefer a neural network or other machine learning approaches to formulate and solve our concerned vaccine distribution problem. However, there would be several downsides to undermine these approaches. Some of the examples can be the requirement of huge historical data, cleansing and consolidation of the dataset, and testing out a broad range of architecture with lengthy training sessions. Needless to say, extensive hyperparameter tuning according to model specifics emerges as another great hassle. Moreover, the potential issues regarding generalization and overfitting of the model will remain questionable for any future unseen test scenarios.%
\vspace{-3mm}
\section{Methodology}\label{methodology}
\vspace{-1mm}
In this section, we initially determine the optimal number and location of vaccine DCs using K-medoids clustering algorithm. Provided with the various locations of possible vaccine DCs, the algorithm determines the optimal number of clusters into which the region can be divided into, in order to effectively distribute the vaccines across the chosen region. On selecting the number of clusters and the cluster heads, we further propose four different vaccine distribution simulation models to optimize vaccine distribution based on two factors namely distance and priority. The clustering algorithm and the simulation models are explained in the subsections below. 
\vspace{-3mm}
\subsection{Selection of Optimal Vaccine Distribution Center}
Our proposed Algorithm \ref{alg:clustering} imitates the core logic of K-medoids clustering technique \cite{park2009simple}. Firstly, the algorithm determines the optimal number of vaccine DCs to be selected from a set of hospitals $\mathcal{\Tilde{H}}$ based on silhouette score analysis \cite{silhouette_score}, as mentioned in line number $1$. We determine the silhouette score for $2$ to $\Tilde{h}_{\Tilde{H}-1}$ number of potential vaccine DCs, where $\mathcal{\Tilde{H}} = \{\Tilde{h}_1, \Tilde{h}_2, \Tilde{h}_3, ... , \Tilde{h}_{\mathcal{\Tilde{H}}}\}$. Then, we select the optimal number of vaccine DCs as $\eta$ that retains the highest silhouette score. As per the line number $2$, we randomly select $\eta$ hospitals as vaccine DCs into $\mathcal{H}$. Later, we initiate the clustering process. Each hospital is assigned to its closest vaccine DC to form $\eta$ clusters, according to line numbers $5-8$. The cluster informations indicating which hospital is associated to which vaccine DC are recorded in $\mathcal{C}$. Next, the algorithm reassigns the vaccine DCs $\mathcal{H}$ to the ones with the minimum total distance to all other hospitals under the same cluster, executed in line numbers $9-16$. We let the algorithm repeat the entire clustering process until vaccine DC assignments do not change. Hence, the termination criteria depends on the stability of the clustering process.
\begin{algorithm}[h!]
\scriptsize
\KwInput{$\mathcal{\Tilde{H}}$: A set of hospitals, $dist$: Squared matrix representing the distances of one hospital to every other hospital}
\KwResult{$\mathcal{H}$: A set of COVID-$19$ vaccine DCs}
     $\eta \gets$ Determine the number of optimal vaccine distribution centers using silhouette score\\
     $\mathcal{H} \gets$ Randomly select $\eta$ hospitals from $\mathcal{\Tilde{H}}$\\
     $\mathcal{C} \gets \varnothing$ \\
     \While{there is no change in $\mathcal{H}$}{
     \ForEach{$\Tilde{h}_{a} \in \mathcal{\Tilde{H}}$}{
     \ForEach{${h}_{b} \in \mathcal{H}$}{
     $h_b \gets$ Find the closest $h_b$ to $\Tilde{h}_a$ using $dist$ matrix
     }
     $\mathcal{C} \gets\mathcal{C} \cup (\Tilde{h}_a, h_b)$\\
     }
     \ForEach{${h}_{b} \in \mathcal{H}$}{
    $temp \gets \varnothing$\\
     \ForEach{$\Tilde{h}_{a} \in \mathcal{\Tilde{H}}$}{
     \If{$(\Tilde{h}_a, h_b)\in \mathcal{C}$}{
     $temp \gets temp \cup \Tilde{h}_a \cup h_{b}$ \\
     }
     }
     $ q \gets \underset{\hat{h}_a \in temp}{\argmin}\hspace{0.16cm} \sum_{h^{*}_{a} \in temp} dist(\hat{h}_a, h^{*}_{a})$\\
     Swap $h_{b}$ with $\Tilde{h}_{q}$ in $\mathcal{C}$\\
     Update $h_{b}$ by $\Tilde{h}_{q}$ in $\mathcal{H}$
     }
     }
\caption{\scriptsize{K-medoids algorithm to choose vaccine DCs from a set of hospitals}}
\label{alg:clustering}
\end{algorithm} 
To summarize, we employ this algorithm with inputs of a set of hospitals/potential vaccine DCs, $\mathcal{\Tilde{H}}$ and a $2$D $dist$ matrix defining the distances of one hospital to every other hospital. The output of the algorithm is the set of optimally selected vaccine DCs $\mathcal{H}$ based on the distance metric. The primary idea is to choose vaccine DCs optimally in a sparse manner to facilitate reachability for people living in any part of the considered region.   
\vspace{-5mm}
\subsection{System Model for Vaccine Distribution}
In this subsection, we proceed by explaining the system model for vaccine distribution. We denote $\mathcal{H} = \{h_1, h_2, h_3, ... , h_{\mathcal{H}}\}$ to be the set of COVID$-19$ vaccine distribution centers selected by Algorithm $1$, where $\mathcal{H} \subseteq \mathcal{\Tilde{H}}$ and $h_i$ is the $i^{th}$ vaccine DC in $\mathcal{H}$. Moreover, we define $\mathcal{U} = \{u_1, u_2, u_3, ..., u_{\mathcal{U}}\}$ as the set of available staff, where every $u_{j} \in \mathcal{U}_i$. We denote $\mathcal{U}_i$ as the available set of staff in vaccine DC $h_{i}$ and $\mathcal{U}_i \subseteq \mathcal{U}$. Subsequently, we can infer that $\mathcal{U} = \cup_{i=1}^{|\mathcal{H}|}\hspace{0.1cm} \mathcal{U}_{i}$. We further assume that $\mathcal{E} = \{e_1, e_2, e_3, ..., e_{\mathcal{E}}\}$ represents the set of people required to be vaccinated, and the $k^{th}$ person is $e_{k}$. In order to specify the priority of people for vaccination purpose, we use the set $\mathcal{P} = \{p_1, p_2, p_3, ..., p_{\mathcal{P}}\}$. Hence, $p_{k}$ defines the priority level of a person $e_k \in \mathcal{E}$, and $|\mathcal{E}|=|\mathcal{P}|$. It is noteworthy that, the higher the priority level, the faster the vaccination service deployment is desired. The distance between a distribution center $h_{i} \in \mathcal{H}$ and a specific person $e_{k} \in \mathcal{E}$ is represented using $\mathcal{D}_{i, k}$. In this research, we consider the solution binary decision variable as $x_{i,j,k} \in \{0, 1\}$. The value of the decision variable $x_{i,j,k}$ is $1$, in case a distribution center $h_{i} \in \mathcal{H}$ allocates a staff $u_j \in \mathcal{U}_i$ to vaccinate a person $e_k \in \mathcal{E}$, otherwise remains $0$.
\vspace{-4mm}
\subsection{Problem Formulation}
In this paper, we formulate the opted vaccine distribution research enigma as a CSP model. The CSP framework includes a set of aforementioned decision variables that should be assigned values in such a way that a set of hard constraints are satisfied. Hard constraints are essential to be satisfied for any model to reach a feasible solution. Suppose, our proposed model iterates over $T = \{t_1, t_2, ..., t_{T}\}$ times to complete vaccination, where each time instance $t_{n} \in T$ refers to per time frame for vaccine deployment decision making. Finally, $\mathcal{N}$ represents the total amount of available vaccine throughout the entire time periods $T$. All of our proposed CSP based models are subject to the following hard constraints that have been translated into integer inequalities, for any time frame $t_{n} \in T$:
\noindent\begin{minipage}{.5\linewidth}
\begin{equation}\label{constraint:1}
     \mathcal{C}1: \sum_{e_k \in \mathcal{E}} x_{i,j,k} \leq 1, \hspace{0.1cm} \forall_{h_i \in \mathcal{H}},\hspace{0.1cm}\forall_{u_j \in \mathcal{U}_i}
\end{equation}
\end{minipage}%
\begin{minipage}{.5\linewidth}
\begin{equation}\label{constraint:2}
     \mathcal{C}2: \sum_{h_i \in \mathcal{H}} \sum_{u_j \in \mathcal{U}_i}  x_{i,j,k} \leq 1, \hspace{0.1cm} \forall_{e_k \in \mathcal{E}} 
\end{equation}
\end{minipage}%
\begin{equation}\label{constraint:3}
     \mathcal{C}3:  x_{i,j,k} \in \{0,1\},  \hspace{0.1cm} \forall_{h_i \in \mathcal{H}},\hspace{0.1cm}\forall_{u_j \in \mathcal{U}_i}\hspace{0.1cm}, \forall_{e_k \in \mathcal{E}} 
\end{equation}
The constraint $\mathcal{C}1$ verifies that every staff from any distribution center can vaccinate at most one person at a time. Thus, for every staff $u_j \in \mathcal{U}_i$ of distribution center $h_i \in \mathcal{H}$, either there is a unique person $e_k \in \mathcal{E}$ assigned for vaccination, or the staff remains unassigned. Then, constraint $\mathcal{C}2$ ensures that every person $e_k \in \mathcal{E}$ is allocated at most one vaccine through a single staff $u_j \in \mathcal{U}_i$ from a unique distribution center $h_i \in \mathcal{H}$. Finally, $\mathcal{C}3$ is a binary constraint representing the value of decision variable to be $1$, in case a staff $u_j \in \mathcal{U}_i$ of distribution center $h_i \in \mathcal{H}$ is assigned to vaccinate a person $e_k \in \mathcal{E}$, otherwise $0$, as mentioned previously.
\begin{equation}\label{constraint:4}
     \mathcal{C}4: \sum_{t_n \in T} \sum_{h_i \in \mathcal{H}} \sum_{u_j \in \mathcal{U}_i} \sum_{e_k \in \mathcal{E}} x_{i,j,k} \leq \mathcal{N}
\end{equation}
The constraint $\mathcal{C}4$ confirms that the total vaccine distribution should not be more than the available vaccine by any means throughout the entire periods $T$ considered for vaccine deployment.
Now, let us assume $\Omega$ be the set of all the feasible solutions that satisfy all hard constraints.
\begin{equation}\label{eq:omega}
      \Omega = \{ x_{i,j,k} \hspace{0.1cm} \vert \hspace{0.1cm} \mathcal{C}1, \mathcal{C}2, \mathcal{C}3, \mathcal{C}4\}
\end{equation}

%\subsection{Soft Constraints}
Apart from hard constraints, our proposed CSP formulation incorporates a set of soft constraints as well. Whilst hard constraints are modeled as inequalities, soft constraints are outlined through expressions intended to be eventually minimized or maximized. Soft constraints are not mandatory for finding a solution, rather highly desirable to improvise the quality of the solutions based on the application domain. The soft constraint $\mathcal{C}5$ strives to maximize the number of overall vaccinated people. The focus of another soft constraint $\mathcal{C}6$ remains to vaccinate the people with higher priority levels beforehand. In other words, this constraint leads to maximize the summation of priorities of all vaccinated people. Subsequently, the soft constraint $\mathcal{C}7$ refers that every people should be vaccinated by staff from the nearest vaccine distribution center. 
\noindent\begin{minipage}{.5\linewidth}
\begin{equation}
    \label{eq:soft_constraint_5}
    \mathcal{C}5: \underset{\Omega}{\max} \sum_{h_i \in \mathcal{H}} \sum_{u_j \in \mathcal{U}_i} \sum_{e_k \in \mathcal{E}}  x_{i,j,k} 
\end{equation}
\end{minipage}%
\begin{minipage}{.5\linewidth}
\begin{equation}
    \label{eq:soft_constraint_6}
    \mathcal{C}6: \underset{\Omega}{\max} \sum_{h_i \in \mathcal{H}} \sum_{u_j \in \mathcal{U}_i} \sum_{e_k \in \mathcal{E}}  x_{i,j,k} \times p_{k}
\end{equation}
\end{minipage}%
\begin{equation}
    \label{eq:soft_constraint_7}
     \mathcal{C}7: \underset{\Omega}{\min} \sum_{h_i \in \mathcal{H}} \sum_{u_j \in \mathcal{U}_i} \sum_{e_k \in \mathcal{E}}  x_{i,j,k} \times \mathcal{D}_{i,k}
\end{equation}
\vspace{-3mm}
\subsection{Variations of Vaccine Distribution Models}
By leveraging different combinations of soft constraints, we propose four different variations of vaccine distribution models: \textbf{a)} Basic - Vaccine Distribution Model (B-VDM), \textbf{b)} Priority based - Vaccine Distribution Model (P-VDM), \textbf{c)} Distance based - Vaccine Distribution Model (D-VDM), and \textbf{d)} Priority in conjunction with Distance based - Vaccine Distribution Model (PD-VDM).

B-VDM is a rudimentary vaccine distribution model that solely concentrates on the soft constraint $\mathcal{C}5$ to maximize the overall vaccine distribution, irrespective of any other factors. A gain co-efficient $\alpha$ has been introduced to the ultimate objective function of the model in Eq. \ref{eq:objective_VM1}.
\begin{equation} \label{eq:objective_VM1}
\mathcal{C}5 \iff \underset{\Omega}{\max} \sum_{h_i \in \mathcal{H}} \sum_{u_j \in \mathcal{U}_i} \sum_{e_k \in \mathcal{E}} \alpha \times x_{i,j,k} 
\end{equation}
P-VDM ensures maximum vaccine distribution among the higher priority groups of people, by reducing soft constraints $\mathcal{C}5$ and $\mathcal{C}6$ into one objective function in Eq. \ref{eq:objective_VM2}. We denote $\beta$ as the gain factor associated to soft constraint $\mathcal{C}6$.
\begin{equation} \label{eq:objective_VM2}
\mathcal{C}5 \wedge \mathcal{C}6 \iff \underset{\Omega}{\max} \sum_{h_i \in \mathcal{H}} \sum_{u_j \in \mathcal{U}_i} \sum_{e_k \in \mathcal{E}} x_{i,j,k} \times (\alpha + \beta \times p_{k})
\end{equation}

Contrarily, D-VDM encourages vaccination of the people located closer to distribution centers. This model can be specifically useful for rural regions, where distribution centers and people are sparsely located, including higher travelling expenses. For this model, we incorporate soft constraints $\mathcal{C}5$ and $\mathcal{C}7$, by multiplying gain coefficients $\alpha$ and $\gamma$, respectively. 
\begin{equation} \label{eq:objective_VM3}
\mathcal{C}5 \wedge \mathcal{C}7 \iff \underset{\Omega}{\max} \sum_{h_i \in \mathcal{H}} \sum_{u_j \in \mathcal{U}_i} \sum_{e_k \in \mathcal{E}} x_{i,j,k} \times (\alpha - \gamma \times \mathcal{D}_{i,k})
\end{equation}

Finally, the PD-VDM merges all the soft constraints simultaneously. Our proposed PD-VDM considers maximization of vaccine distribution in priority groups and minimization of distance factored in transportation expenditure, collaboratively. Furthermore, $\alpha$, $\beta$, and $\gamma$ have been introduced as gain factors to equilibrate the combination of soft constraints and then presented as a multi-objective function in the Eq. \ref{eq:objective_VM4}. Hence, this model can optimize priority and distance concerns conjointly based on the adapted values of gain factors. 

\begin{equation} \label{eq:objective_VM4}
\mathcal{C}5 \wedge \mathcal{C}6 \wedge \mathcal{C}7 \iff \underset{\Omega}{\max} \sum_{h_i \in \mathcal{H}} \sum_{u_j \in \mathcal{U}_i} \sum_{e_k \in \mathcal{E}} x_{i,j,k} \times (\alpha + \beta \times p_{k} - \gamma \times \mathcal{D}_{i,k})
\end{equation}
The gain parameters of all the proposed models can be tuned to balance or incline towards more focused convergent vaccine distribution solutions. For instance, $\alpha$, $\beta$, and $\gamma$ are individually responsible for maximum distribution, maximization of priorities, and minimization of distance focused vaccine distribution solutions, respectively. Moreover, the policymakers can exploit these models and adjust gain parameters according to the region specifics and domain knowledge of vaccine distribution centers and population density. The regulation of these gain factors can aid to analyze and figure out the applicability of our various proposed models relying on different contextual targets, environment settings, and demand-supply gaps.
\vspace{-3mm}
\section{Case Study - Chennai, India}\label{casestudy}
We demonstrate the proposed models using real-world data obtained from one of the popular cities in the southern part of India - Chennai. As listed by Rachaniotis et al. \cite{rachaniotis2012deterministic} most of the articles in literature make various assumptions to demonstrate the performance of their models, Similarly, we make a few reasonable assumptions to accommodate the lack of crucial data required to implement the model. Various input parameters of the models and their method of estimation or assumptions made to reach the decisions are described below.
\vspace{-3mm}
\subsection{Clustering phase}
    \begin{itemize}
         \item The entire city is divided into 15 zones for administrative purposes and we assume the distribution of vaccines is carried out based on these 15 zones as well.
         \item To determine the vaccine distribution centers, we assume that the vaccines will be distributed from hospitals or primary health centers. While there are approximately 800 hospitals in Chennai, based on `on-the-ground' knowledge, we select 45 hospitals (3 hospitals per zone) to enable us to determine the distance between the hospitals. The selected hospitals are classified as private (PVT) and public funded (PUB) based on their administration. At least one publicly funded hospital or primary health center is chosen per zone. 
         \item A 45 x 45 distance matrix is constructed with each row and column representing the hospitals. The cells are populated with the geographic distance obtained from Google Maps\footnote{https://www.google.com/maps/}. Using these distance measures, the optimal vaccine DCs are chosen by implementing k-medoids clustering algorithm described in Section 3.1. 
    \end{itemize}
\vspace{-3mm}
\subsection{Vaccine-Distribution Phase}
    \begin{itemize}
    \item \textbf{Total population to be vaccinated ($\mathcal{E}$):} As per the census records collected in 2011, Chennai has a population of 4.6 million distributed across an area of 175 km\textsuperscript{2} with a population density of 26,553 persons/km\textsuperscript{2}. Based on the growth in the overall population of India, we estimate the current population of Chennai to be 5,128,728 with a population density of 29,307/km\textsuperscript{2}. 
    % \begin{equation}
    %       P_n = \big( \frac{r}{100} \times P_{n-1}\big) + P_{n-1}  
    % \end{equation}
    % \vspace{-5mm}
    % \begin{figure}[h]
    %     \centering
    %     \begin{tabular}{@{}>{$}l<{$}l@{}}
    %         where:\\
    %         P_n     & Total population during the current year\\
    %         P_{n-1} & Total population during the previous year\\
    %         r       & rate of increase of overall population in India\\
    %     \end{tabular}
    % \end{figure}
    \item \textbf{Set of Vacccine Distribution centers ($\mathcal{H}$):} The optimal vaccine DCs are chosen from using the clustering phase of the model using silhouette width such that the chosen DCs are evenly spread across the entire district. 
    \item \textbf{Staff for vaccination per DC ($\mathcal{U}$):} Based on the capacity of the hospital in terms of facilities, workforce, etc. the hospitals are classified as ‘SMALL’, ‘MED’, and ‘LARGE’. We assume that small, medium, and large hospitals allocate 5, 20, and 40 health-care workers for vaccination purposes respectively.
    \item \textbf{Priority levels ($\mathcal{P}$):} While the priority levels can be decided by the authorities based on wide variety of parameters, for our simulation we assume that the priority depends on the age of the individual such that the elderly people are vaccinated earlier. The distribution of population across various age groups is calculated based on the age-wise distribution calculated during the census 2011. Table \ref{tab:agetable} shows the distribution of the population across six priority groups.
    \begin{table}[h]
    \centering
    \begin{tabular}{|p{3.25cm}|p{1cm}|p{1cm}|p{1cm}|p{1cm}|p{1cm}|p{1cm}|}
    \hline
    \textbf{Age group}          & 0-9   & 10-19 & 20-54 & 55-64 & 65-74 & 75+  \\ \hline
    \textbf{\% of distribution} & 14.02 & 15.34 & 56.38 & 7.87  & 4.09  & 2.31 \\ \hline
    \end{tabular}
    \caption{Distribution of the population of Chennai across 6 age groups}
    \label{tab:agetable}
    \vspace{-8mm}
    \end{table}
    \item \textbf{Time per vaccination ($t$):} Based on the data provided during the recent trial dry run carried out in India, the government estimates to carry out 25 vaccinations in 2 hours span, we approximate that the time taken for the vaccination of one person to be 5 minutes.
    \item \textbf{Time per vaccination ($\mathcal{N}$):} Keeping in mind the deficit in the supply of vaccines in the early stages of vaccination we assume that only vaccine doses for 50\% of the total population is available currently. However, the number can be increased based on increase in production \footnote{https://www.cnn.com/2020/12/18/asia/india-coronavirus-vaccine-intl-hnk/index.html}
    \end{itemize}
    These described parameters can be tuned by the decision making authorities to accommodate the distribution at the area under consideration. For all the experimental settings, we consider the gain factors $\alpha$, $\beta$, and $\gamma$ as $\frac{1}{4} \times |\epsilon|$, $\frac{1}{4} \times \frac{1}{|\mathcal{P}|} \times  |\epsilon|$, and $1$ respectively. Yet, as mentioned earlier, these gain factors can be explored and set according to the solutions desired by policy makers and decision making authorities.
\vspace{-3mm}
\section{Results and Discussion}\label{results}
\vspace{-1mm}
We demonstrate the proposed models in two scenarios with randomly generated data and two scenarios with real world data from Chennai. We discuss in detail the inferences from each scenario in this section. In each scenario we compare the four models and highlight how PD-VDM optimizes the two parameters taken into consideration - priority and distance.
\vspace{-3mm}
\subsection{Random-simulation} 
\begin{figure*}
        \centering
        \begin{subfigure}[b]{0.30\textwidth}
            \centering
            \includegraphics[width=\textwidth]{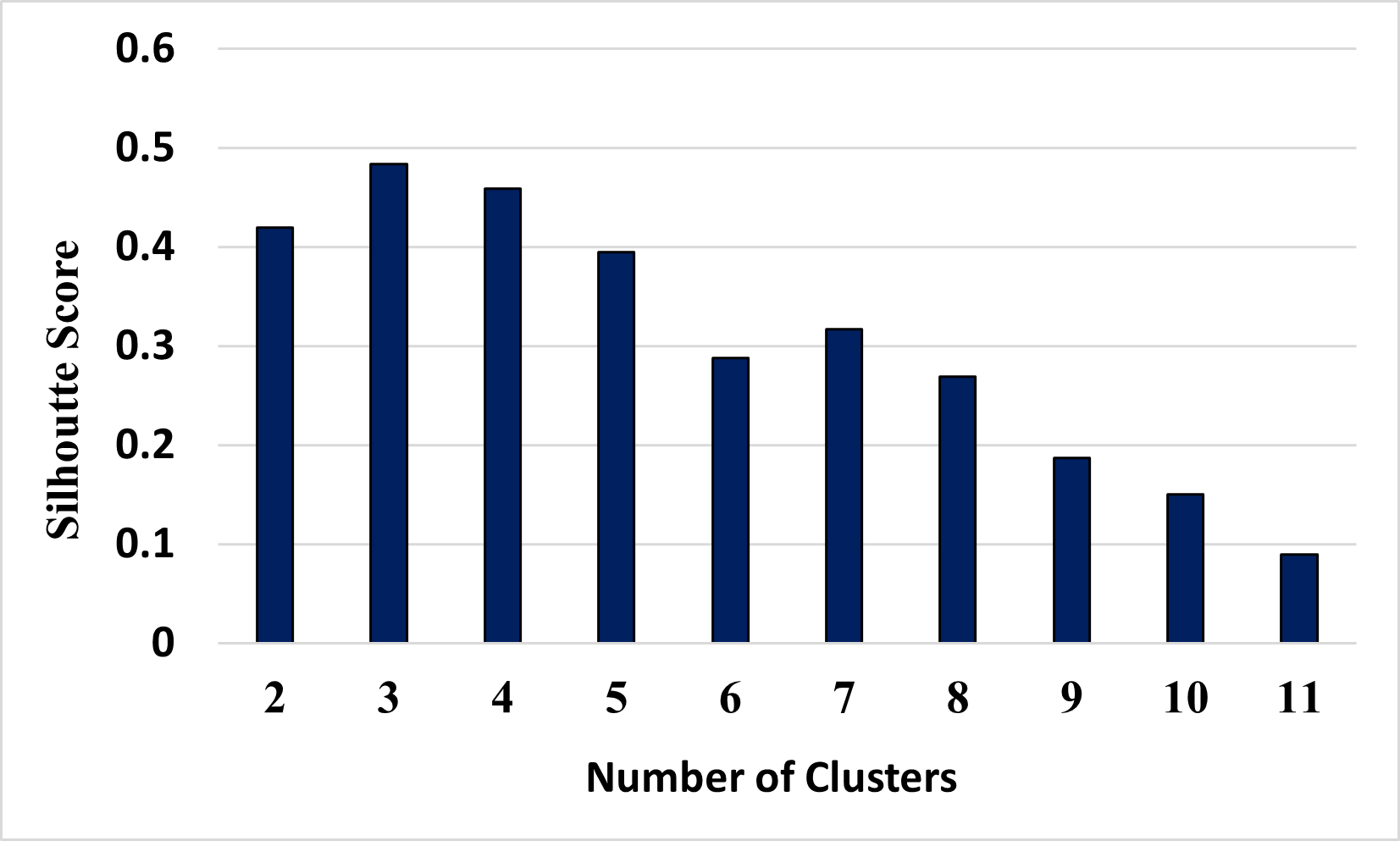}
            \caption[Random simulation]
            {{\small Random simulation}}    
            \label{RS}
        \end{subfigure}
        \hfill
        \begin{subfigure}[b]{0.69\textwidth}  
            \centering 
            \includegraphics[width=\textwidth]{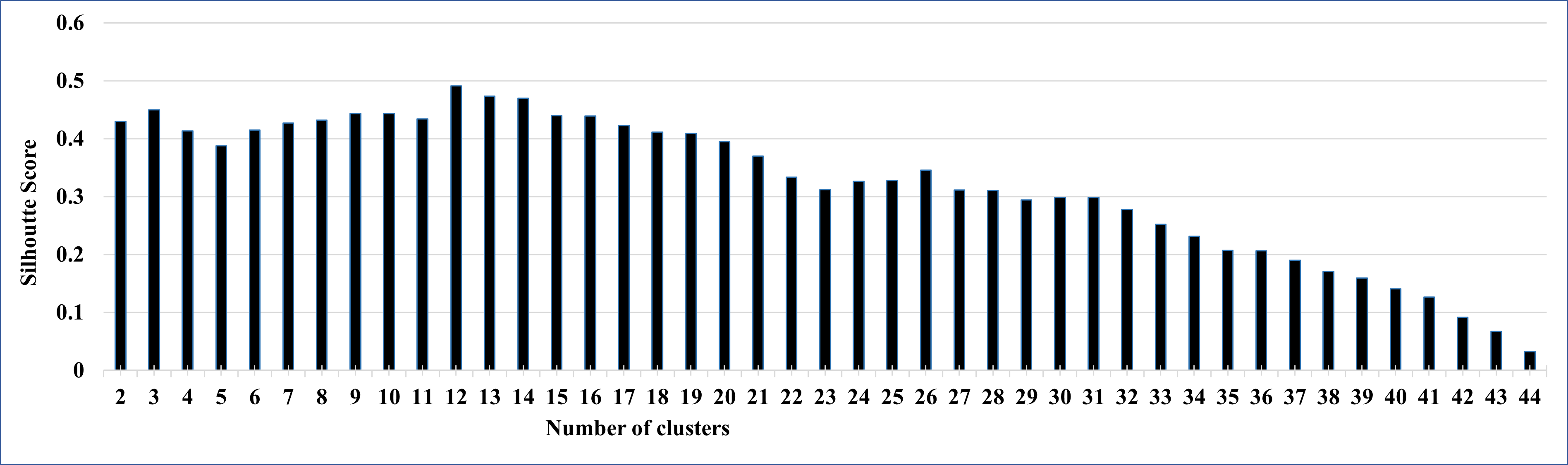}
            \caption[Case Study]%
            {{\small Case Study }}    
            \label{CS}
        \end{subfigure}
         \caption[Selection of optimum number of DCs based on Silhouette width]
        {\small Distribution of vaccine across various priority groups} 
        \label{fig:silhouette}
    \end{figure*}

\vspace{-1mm}
For the two random scenarios we consider that there are 12 vaccine DCs in total and based on the silhouette width measures as shown in Fig. \ref{RS} we select three DCs optimally distributed across the area. We assume the total population size ($\mathcal{E}$) to be 200 and the total number of vaccines available ($\mathcal{N}$) to be 85. Of the three chosen DCs we assume that each has a capacity ($\mathcal{U}$) of 15, 30, 45. The model considers five different different priority levels ($\mathcal{P}$) with 1 being the least and 5 being the most. The population is distributed among these priority levels as shown in Table \ref{tab:table_rand}. 
  \begin{table}[h!]
    \centering
    \begin{tabular}{{|p{4.5cm}|p{1cm}|p{1cm}|p{1cm}|p{1cm}|p{1cm}|}}
    \hline
    \textbf{Priority group} & 1  & 2  & 3  & 4  & 5  \\ \hline
    \textbf{Raw Count}      & 43 & 35 & 50 & 45 & 27 \\ \hline
    \end{tabular}
    \caption{Distribution of population across priority groups in random simulation.}
    \label{tab:table_rand}
    \vspace{-6mm}
    \end{table}
To demonstrate the impact of the population distribution parameter, we run the simulation model under two different distributions namely, 
%(a) Random-case -1 (RC1): Uniform random distribution (b) Random-case -2 (RC2): Poisson like distribution where the population is dense at some regions and spares at others.
\begin{itemize}
    \item Random-case -1 (RC-1): Uniform random distribution
    \item Random-case -2 (RC-2): Poisson like distribution where the population is dense at some regions and spares at others.
\end{itemize}
\vspace{-2mm}
\begin{figure*}
        \centering
        \begin{subfigure}[b]{0.24\textwidth}
            \centering
            \includegraphics[width=\textwidth]{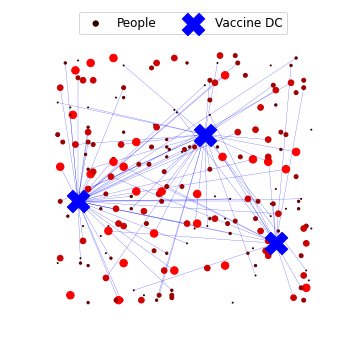}
            \caption[RC1 B-VDM]%
            {{\small RC1 B-VDM}}    
            \label{RC1B}
        \end{subfigure}
        \hspace{1mm}
        \begin{subfigure}[b]{0.24\textwidth}  
            \centering 
            \includegraphics[width=\textwidth]{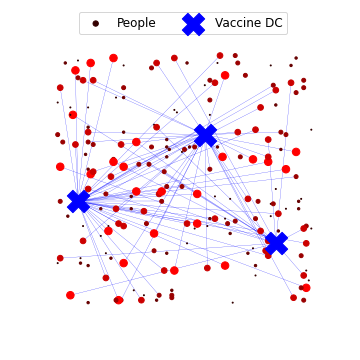}
            \caption[RC1 P-VDM]%
            {{\small RC1 P-VDM}}    
            \label{RC1P}
        \end{subfigure}
          \begin{subfigure}[b]{0.24\textwidth}
            \centering
            \includegraphics[width=\textwidth]{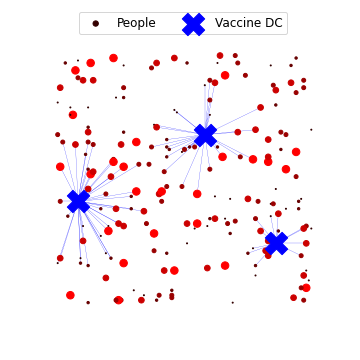}
            \caption[RC1 D-VDM]%
            {{\small RC1 D-VDM}}    
            \label{RC1D}
        \end{subfigure}
        \hfill
        \begin{subfigure}[b]{0.24\textwidth}  
            \centering 
            \includegraphics[width=\textwidth]{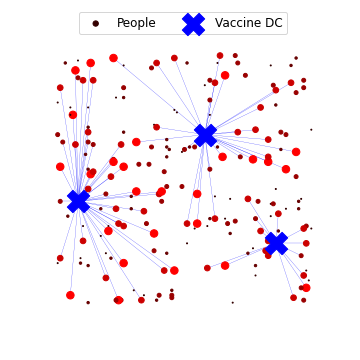}
            \caption[RC1 PD-VDM]%
            {{\small RC1 PD-VDM}}    
            \label{RC1PD}
        \end{subfigure}
        \vskip\baselineskip
        \begin{subfigure}[b]{0.24\textwidth}
            \centering
            \includegraphics[width=\textwidth]{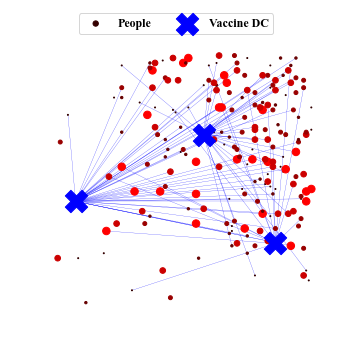}
            \caption[RC2 B-VDM]%
            {{\small RC2 B-VDM}}    
            \label{RC2B}
        \end{subfigure}
        \hfill
        \begin{subfigure}[b]{0.24\textwidth}  
            \centering 
            \includegraphics[width=\textwidth]{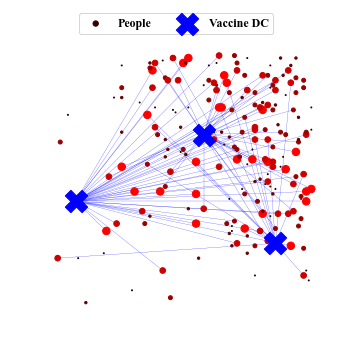}
            \caption[RC2 P-VDM]%
            {{\small RC2 P-VDM}}    
            \label{RC2P}
        \end{subfigure}
          \begin{subfigure}[b]{0.24\textwidth}
            \centering
            \includegraphics[width=\textwidth]{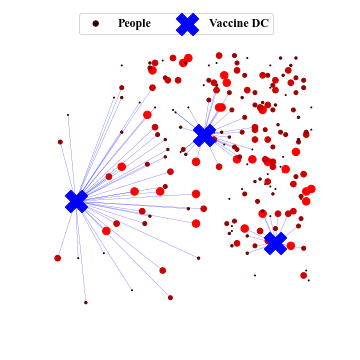}
            \caption[RC2 D-VDM]
            {{\small RC2 D-VDM}}    
            \label{RC2D}
        \end{subfigure}
        \hfill
        \begin{subfigure}[b]{0.24\textwidth}  
            \centering 
            \includegraphics[width=\textwidth]{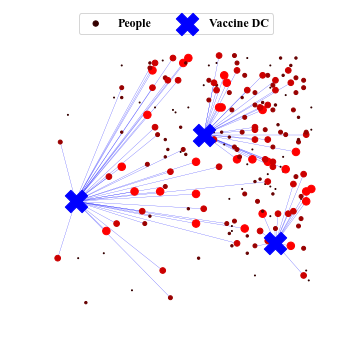}
            \caption[RC2 PD-VDM]%
            {{\small RC2 PD-VDM}}    
            \label{RC2PD}
        \end{subfigure}
        \caption[ Snapshot of the simulation of the models in the random case studies]
        {\small Snapshot of the simulation of the models in the random case studies} 
        \label{fig: snapshot}
    \end{figure*}
\begin{figure*}
        \centering
        \begin{subfigure}[b]{0.50\textwidth}
            \centering
            \includegraphics[width=\textwidth]{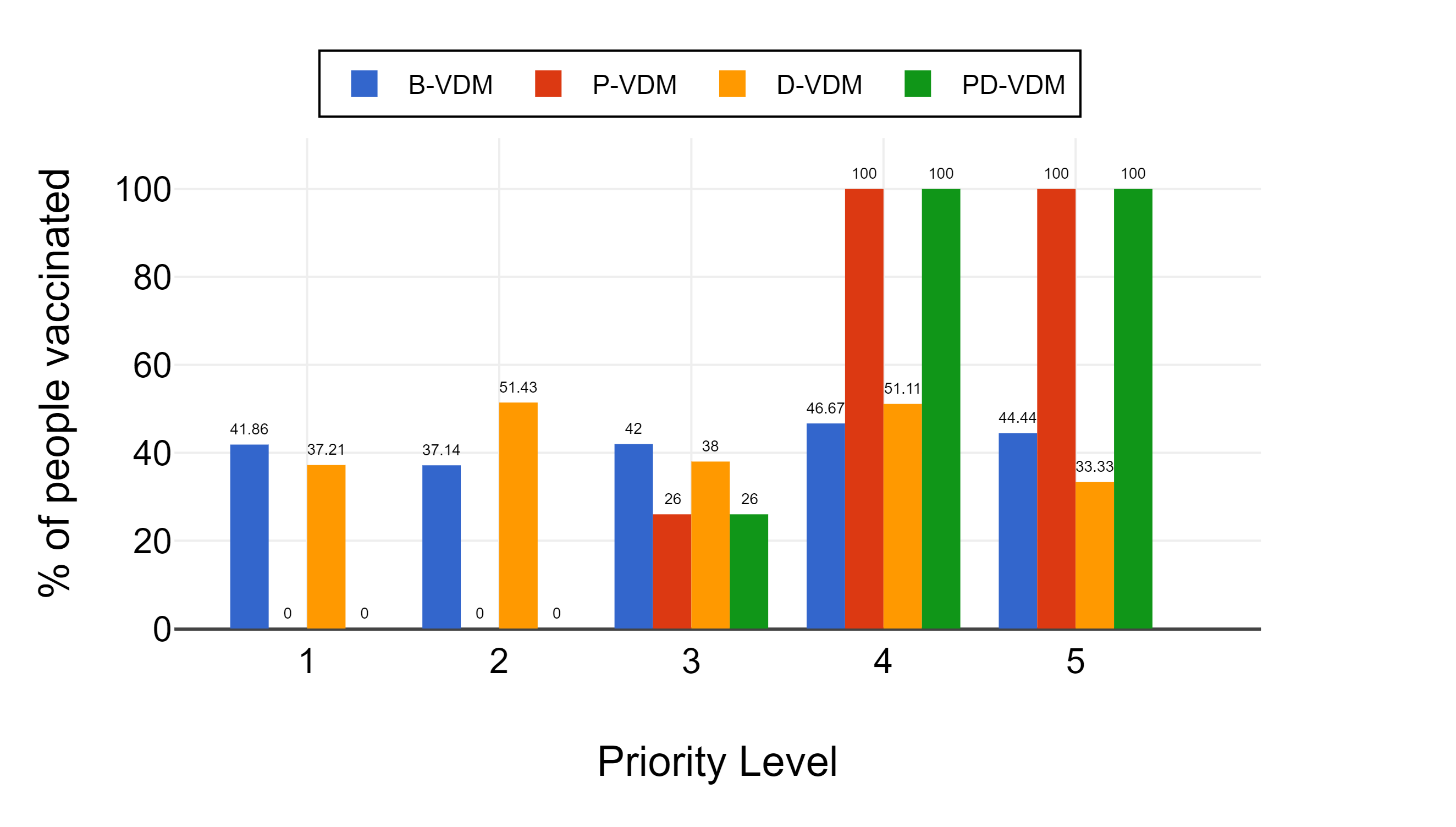}
            \caption[RC1]%
            {{\small RC1}}    
            \label{Prior-RC1}
        \end{subfigure}
        \hfill
        \begin{subfigure}[b]{0.49\textwidth}  
            \centering 
            \includegraphics[width=\textwidth]{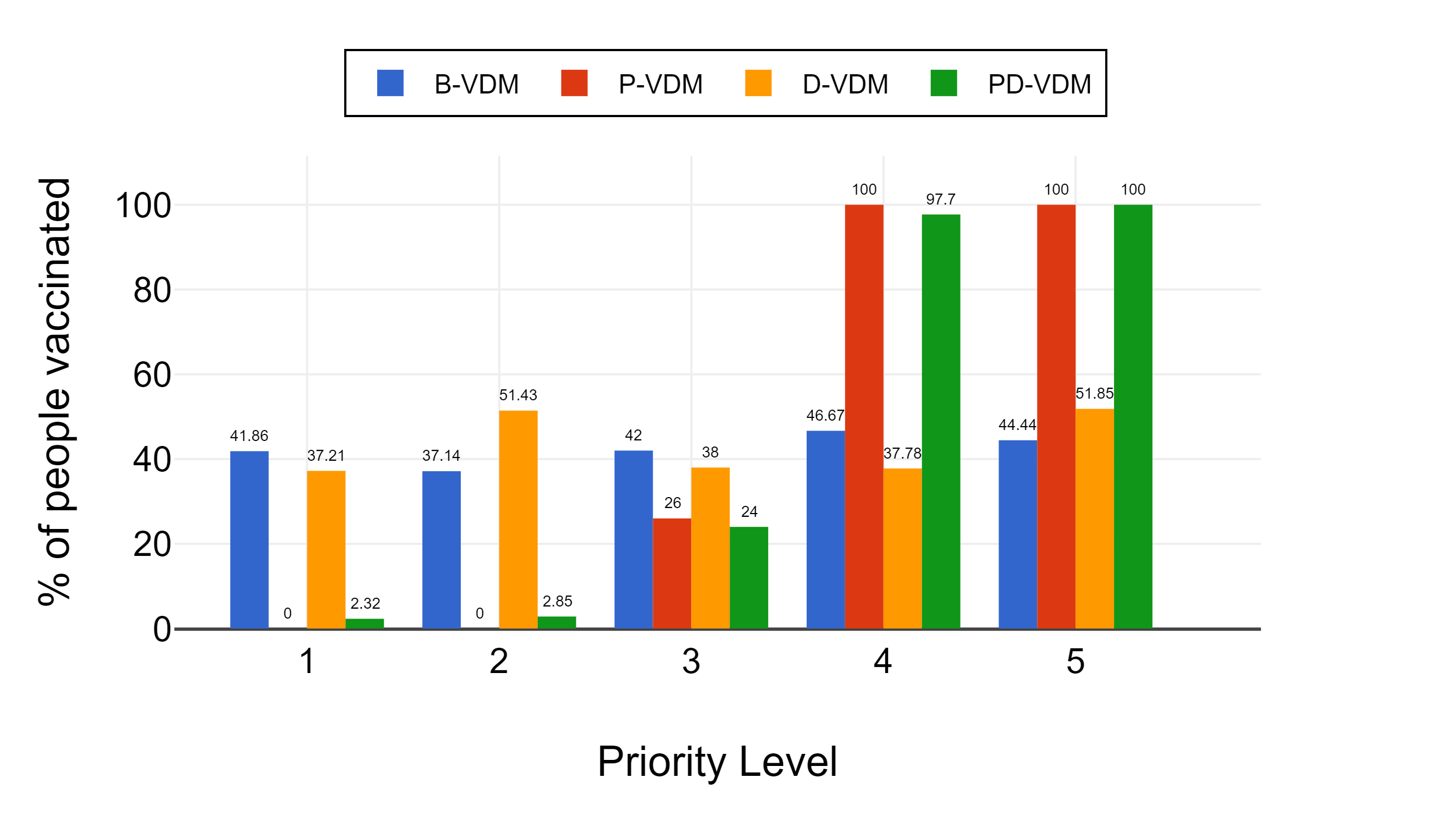}
            \caption[RC2]%
            {{\small RC2}}    
            \label{Prior-RC2}
        \end{subfigure}
        \begin{subfigure}[b]{0.495\textwidth}
            \centering
            \includegraphics[width=\textwidth]{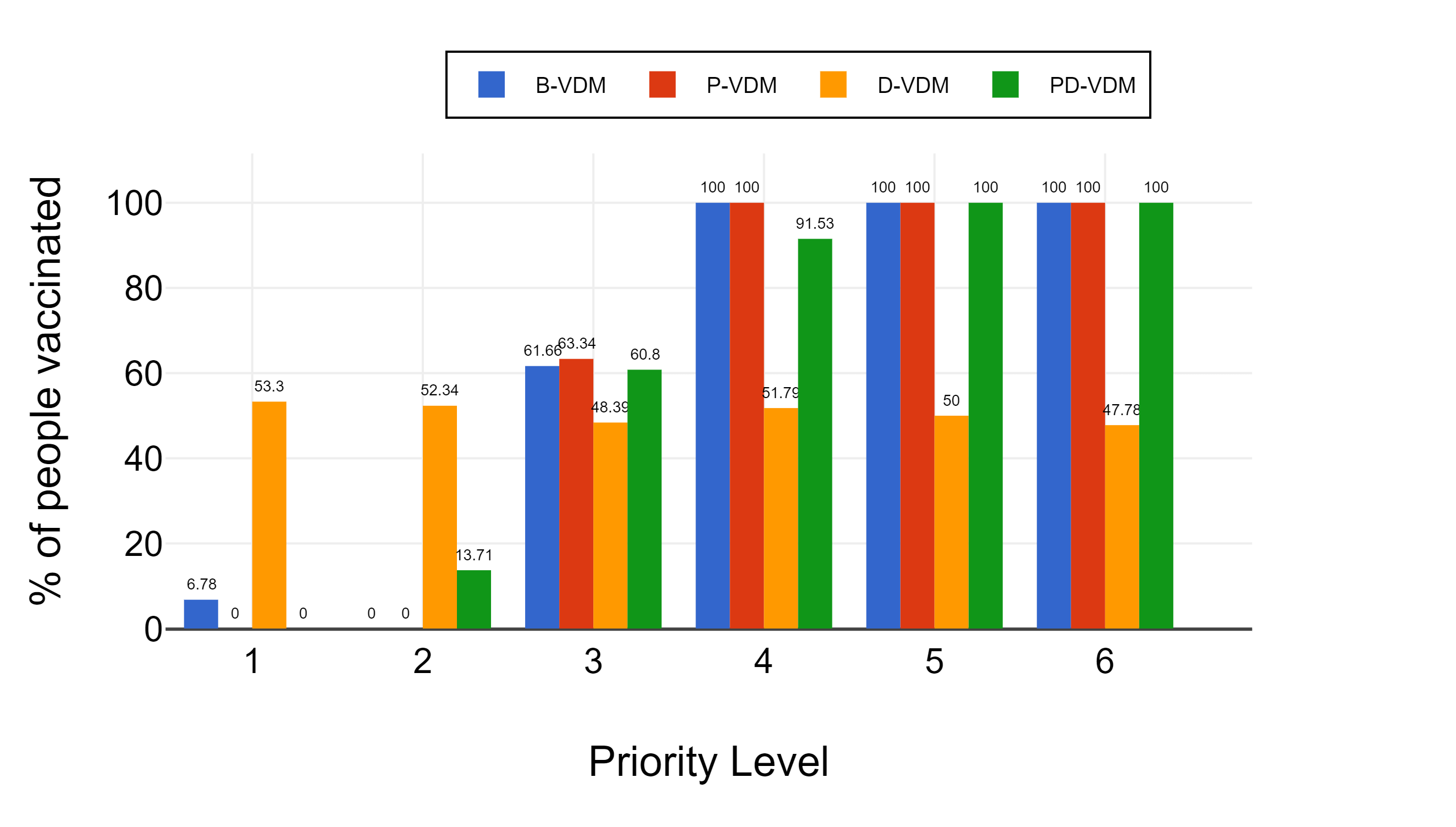}
            \caption[CS1]%
            {{\small CS1}}    
            \label{Prior-CS1}
        \end{subfigure}
        \hfill
        \begin{subfigure}[b]{0.495\textwidth}  
            \centering 
            \includegraphics[width=\textwidth]{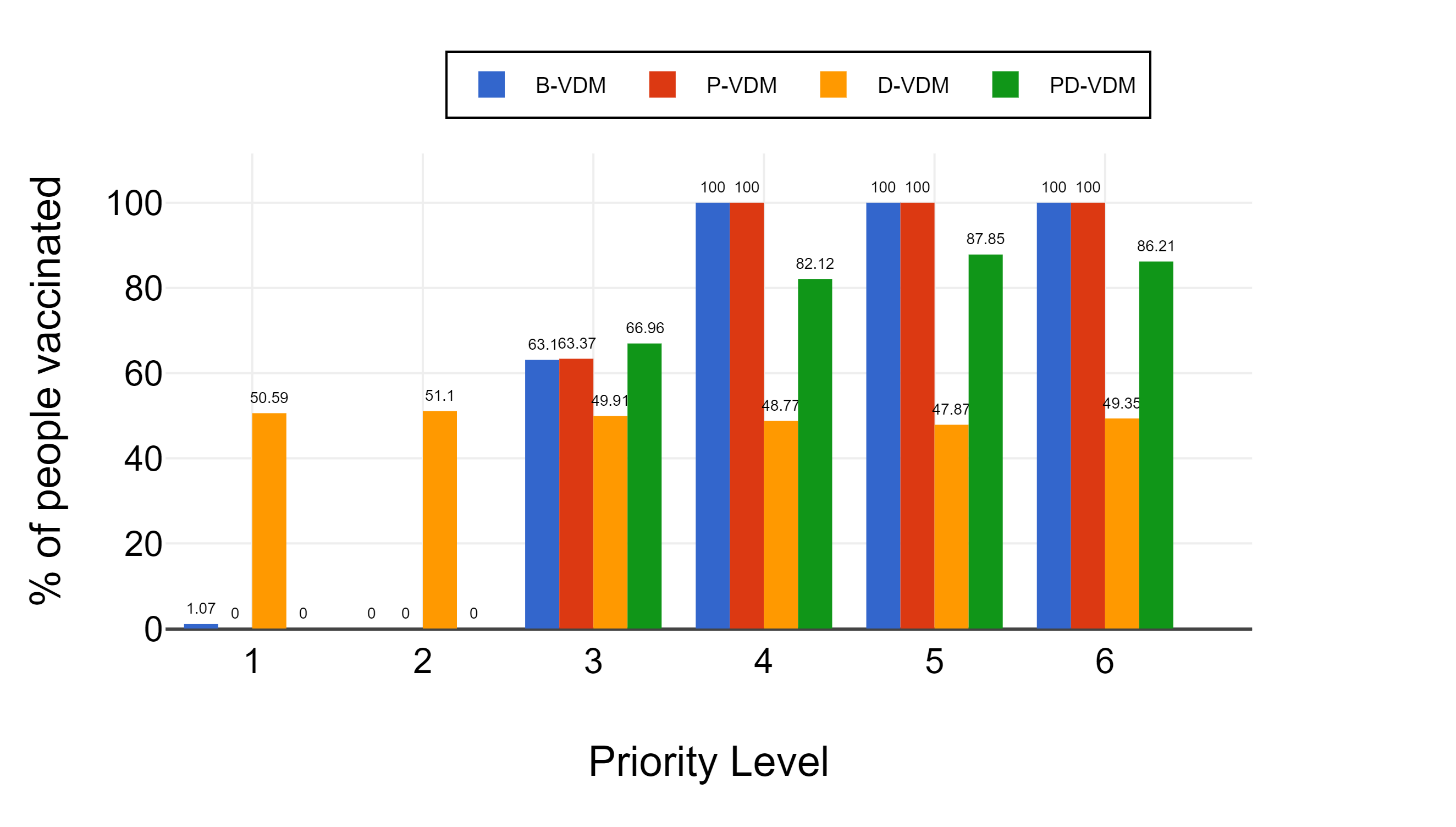}
            \caption[CS2]%
            {{\small CS2}}    
            \label{Prior-CS2}
        \end{subfigure}
        \caption[Distribution of vaccine across various priority groups]
        {\small Distribution of vaccine across various priority groups} 
        \label{fig:distribution}
    \end{figure*}
\par
\begin{figure*}
        \centering
        \begin{subfigure}[b]{0.495\textwidth}
            \centering
            \includegraphics[width=\textwidth]{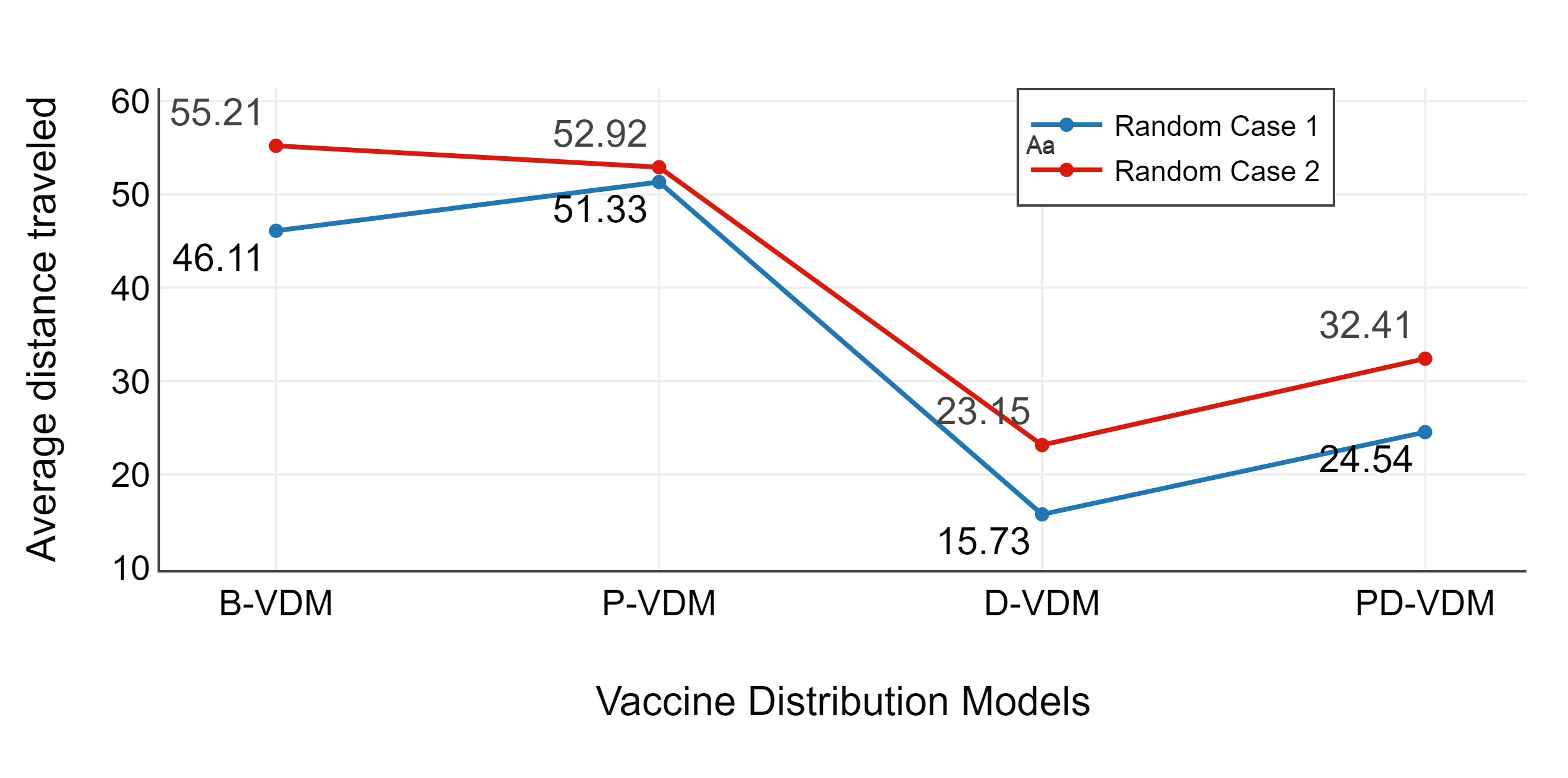}
            \caption[ RC1 and RC2]%
            {{\small RC1 and RC2}}    
            \label{dist-rand}
        \end{subfigure}
        \begin{subfigure}[b]{0.495\textwidth}  
            \centering 
            \includegraphics[width=\textwidth]{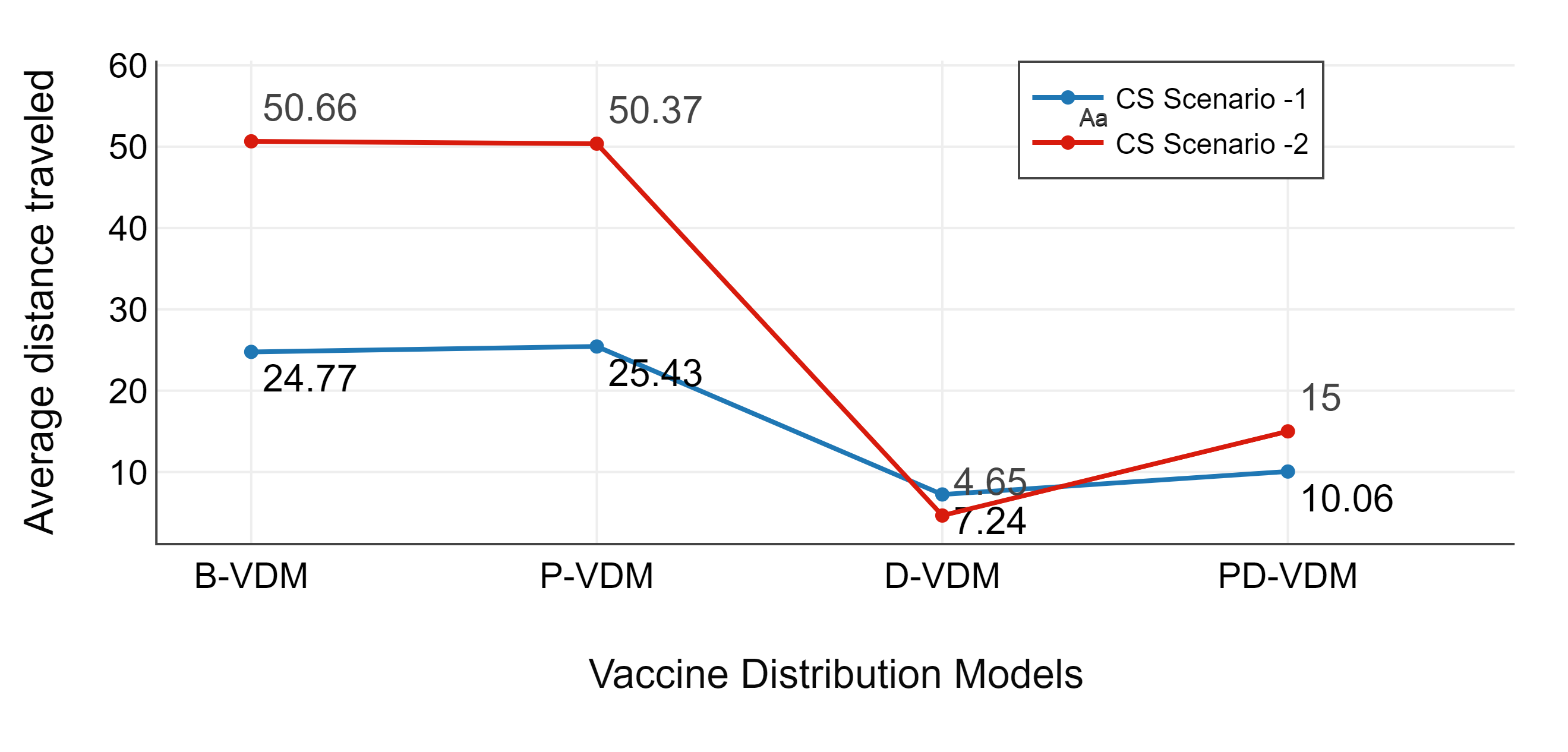}
            \caption[ CS1 and CS2]%
            {{\small CS1 and CS2}}    
            \label{dist-case}
        \end{subfigure}
         \caption[Distribution of vaccine across various priority groups]
        {\small Distribution of vaccine across various priority groups} 
        \label{fig:distribution}
    \end{figure*}

 The distribution of vaccines at any time instance $t_{n}$ by all four models, for both random cases is depicted in Fig. \ref{fig: snapshot}. The percentage of individuals vaccinated in each priority group in random case 1 and random case 2 are depicted in Fig.\ref{Prior-RC1} and Fig. \ref{Prior-RC2} respectively. The B-VDM vaccinates 41.86\%, 37.14\%, 42\%, 46.67\%, and 44.44\%  of individuals across priorities 1 to 5 respectively, in both RC1 and RC2. We can see that 55.56\% of the individuals from the highest priority group are left out. The D-VDM optimizes only the distance parameter and vaccinates 37.21\%, 51.43\%, 38\%, 51.11\% and 33.33\% of individuals across priorities 1 to 5 respectively in RC1 and almost the same results in RC2. Again we can notice, that a greater percentage of the individuals in the highest priority group are not vaccinated. While both P-VDM and PD-VDM attempt to vaccinate 100\% of the high priority individuals in both RC1 and RC2, on studying the average distance travelled by each individual of the population as depicted in Fig. \ref{dist-rand}, we can clearly identify that the PD-VDM reduces the distance parameter by more than 40\% in both RC1 and RC2 . We can also see that the `distribution of population' parameter does not impact the performance of the models and PD-VDM is the most efficient across both Uniform and Poisson distribution. Although there is a slight increase in the average distance travelled in the second distribution the PD-VDM still achieves the most optimal results.
\vspace{-8mm}
\vspace{-14mm}
\subsection{Case-Study simulation}  
In both the case-study scenarios we consider that the vaccination process continues for five hours each day and as mentioned in Section \ref{casestudy} each vaccination takes ($t_{n}$) 5 mins. Hence there are 60 vaccinations carried out by each health care worker for any given day. We compile and present the results of the vaccination process at the end of any given day. Although in the initial stages of the vaccination it is likely that individuals from one priority group will only be vaccinated, as time progresses people across various priority groups will need to be considered. Hence the population considered by the model for each day is taken as a stratified sample from the age-wise distribution of the overall population of Chennai provided in Table \ref{tab:agetable}. Initially we identify the optimal number of DCs needed for effectively serving the district of Chennai based on the distance between the hospitals considered. On analyzing the silhouette width of various `number of clusters' as depicted in Fig. \ref{CS} we present two different scenarios namely,
\begin{itemize}
    \item Case study scenario -1 (CS1): Three optimal DCs
    \item Case study scenario -2 (CS2): Twelve optimal DCs
\end{itemize}
\vspace{-6mm}
\subsubsection{Case study scenario -1:}For this scenario, we consider 3 optimal vaccine DCs ($\mathcal{H}$) with capacities ($\mathcal{U}$) of 5, 20, 40 since the chosen DCs fall under `SMALL', `MED' and `LARGE' respectively. Based on these factors the total population to be vaccinated ($\mathcal{E}$) in any given data is estimated to be 3,900. As mentioned in Section \ref{casestudy} we assume, that the total vaccines ($\mathcal{N}$) available to be half the total population which in this case sums upto 1,950. The entire population falls under six priority groups ($\mathcal{P}$) based on their age. The actual number of people in each priority group is provided in Table \ref{tab:dist_case_study}. 
\begin{table*}[t!]
\centering
\begin{tabular}{|p{4.5cm}|p{1cm}|p{1cm}|p{1cm}|p{1cm}|p{1cm}|p{1cm}|p{1cm}}
\hline
\textbf{Priority group}        & \textbf{1} & \textbf{2} & \textbf{3} & \textbf{4} & \textbf{5} & \textbf{6} \\ \hline
\textbf{Case-study scenario 1} & 546        & 598        & 2199       & 307        & 160        & 90         \\ \hline
\textbf{Case-study scenario 2} & 2817       & 3082       & 11331      & 1583       & 823        & 464        \\ \hline
\end{tabular}
\caption{Distribution of population across six priority groups for Case study scenarios}
\label{tab:dist_case_study}
\vspace{-10mm}
\end{table*}
\subsubsection{Case study scenario -2:} For this scenario, we consider 12 optimal vaccine DCs ($\mathcal{H}$) and among these 12 chosen DCs there are 3, 2 and 7 DCs with a staff capacity ($\mathcal{U}$) of 5 (`SMALL'), 20 (`MED'), and 40 (`LARGE') respectively. Hence the total population ($\mathcal{E}$) that can be vaccinated at any given 5 hour day is 20100. Similar to scenario 1 we assume that the total vaccines ($\mathcal{N}$) available to be half the total population which amounts to 10,050 and that the population is distributed across six different priority groups ($\mathcal{P}$) based on their age as shown in Table \ref{tab:dist_case_study}.

The vaccination percentage across the priority groups for both case study scenarios are provided in Fig. \ref{Prior-CS1} and Fig. \ref{Prior-CS2} respectively. Unlike the random scenarios, it is interesting to note that in both CS1 and CS2, the B-VDM though not optimized to satisfy a specific hard constraint, it produces results that are almost identical to P-VDM vaccinating 100\% of the highest priority group individuals. This can be attributed to the similarity in the effect of $\alpha$ and $\beta$ gain paramters in both these models in CS1 and CS2. Though the PD-VDM model vaccinates less than 90\% of the individuals in the three highest priority groups we can clearly see that it significantly reduces the `average distance travelled' by an individual in the population by more than 70\%, which makes it more efficient than all the other models. While D-VDM achieves the least `average distance travelled' value it sacrifices vaccinating almost 50\% of the high priority groups. Thus, we demonstrate how PD-VDM efficiently distributes the available vaccines by modifying various parameters like the distribution of population, total population, total number of available vaccines, the number of vaccine DCs and the capacity of each DCs. The model provides flexibility for the decision-making authorities of any given demographic to optimize the distribution of vaccine in the desired region.
\vspace{-3mm}
\section{Conclusion and Future Work}\label{conclusion}
\vspace{-2mm}
In this paper, we propose an optimization model (PD-VDM) based on Constraint Satisfaction Programming framework to find the most effective way to distribute vaccines in a given demographic region, in terms of distance and a priority (age, exposure, vulnerability, etc). We compare the efficiency of the model with three other models which take into consideration either one or none of the two optimization constraints. While this model can be adapted across a wide variety of scenarios as demonstrated in our case studies, it is essential to understand that due to resource constraints we have demonstrated only two of the many available constraints. Expanding the scope of the model to allow optimizing a wide variety of parameters can be carried out in future research, along with an attempt to replace some of the assumptions made in our model with real world data. 
% %
% % ---- Bibliography ----
% %
% % BibTeX users should specify bibliography style 'splncs04'.
% % References will then be sorted and formatted in the correct style.
% %

\bibliographystyle{splncs04}
\bibliography{references}

\begin{thebibliography}{10}
\providecommand{\url}[1]{\texttt{#1}}
\providecommand{\urlprefix}{URL }
\providecommand{\doi}[1]{https://doi.org/#1}

\bibitem{aaby2006montgomery}
Aaby, K., Herrmann, J.W., Jordan, C.S., Treadwell, M., Wood, K.: Montgomery
  county’s public health service uses operations research to plan emergency
  mass dispensing and vaccination clinics. Interfaces  \textbf{36}(6),
  569--579 (2006)

\bibitem{ali2020covid}
Ali, I., Alharbi, O.M.: {COVID}-19: Disease, management, treatment, and social
  impact. Science of The Total Environment  \textbf{728},  138861 (2020)

\bibitem{begen2016supply}
Begen, M.A., Pun, H., Yan, X.: Supply and demand uncertainty reduction efforts
  and cost comparison. International Journal of Production Economics
  \textbf{100}(180),  125--134 (2016)

\bibitem{CSP-Survery}
Bordeaux, L., Hamadi, Y., Zhang, L.: Propositional satisfiability and
  constraint programming: A comparative survey. ACM Computing Surveys (CSUR)
  \textbf{38}(4),  12--es (2006)

\bibitem{brandeau2005allocating}
Brandeau, M.L.: Allocating resources to control infectious diseases. In:
  Operations Research and Health Care, vol.~70, pp. 443--464. Springer (2005)

\bibitem{dong2020interactive}
Dong, E., Du, H., Gardner, L.: An interactive web-based dashboard to track
  {COVID}-19 in real time. The Lancet infectious diseases  \textbf{20}(5),
  533--534 (2020)

\bibitem{duijzer2018literature}
Duijzer, L.E., van Jaarsveld, W., Dekker, R.: Literature review: The vaccine
  supply chain. European Journal of Operational Research  \textbf{268}(1),
  174--192 (2018)

\bibitem{halper2011mobile}
Halper, R., Raghavan, S.: The mobile facility routing problem. Transportation
  Science  \textbf{45}(3),  413--434 (2011)

\bibitem{Huang2017}
Huang, H.C., Singh, B., Morton, D., Johnson, G., Clements, B., Meyers, L.:
  Equalizing access to pandemic influenza vaccines through optimal allocation
  to public health distribution points. PLOS ONE  \textbf{12},  e0182720 (08
  2017). \doi{10.1371/journal.pone.0182720}

\bibitem{kaur2020covid}
Kaur, S.P., Gupta, V.: {COVID}-19 {V}accine: A comprehensive status report.
  Virus {R}esearch  \textbf{288},  198114 (2020)

\bibitem{lee2013advancing}
Lee, E.K., Pietz, F., Benecke, B., Mason, J., Burel, G.: Advancing public
  health and medical preparedness with operations research. Interfaces
  \textbf{43}(1),  79--98 (2013)

\bibitem{mahase2020covid}
Mahase, E.: {COVID}-19: Moderna vaccine is nearly 95\% effective, trial
  involving high risk and elderly people shows. BMJ: British Medical Journal
  (Online)  \textbf{371} (2020)

\bibitem{nicola2020socio}
Nicola, M., Alsafi, Z., Sohrabi, C., Kerwan, A., Al-Jabir, A., Iosifidis, C.,
  Agha, M., Agha, R.: The socio-economic implications of the coronavirus and
  {COVID}-19 pandemic: A review. International Journal of Surgery  \textbf{78},
   185--193 (2020)

\bibitem{park2009simple}
Park, H.S., Jun, C.H.: A simple and fast algorithm for {K}-medoids clustering.
  Expert systems with applications  \textbf{36}(2),  3336--3341 (2009)

\bibitem{polack2020safety}
Polack, F.P., Thomas, S.J., Kitchin, N., Absalon, J., Gurtman, A., Lockhart,
  S., Perez, J.L., P{\'e}rez~Marc, G., Moreira, E.D., Zerbini, C., et~al.:
  Safety and efficacy of the {BNT}162b2 m{RNA} {COVID}-19 vaccine. New England
  Journal of Medicine  \textbf{383},  2603--2615 (2020)

\bibitem{rachaniotis2012deterministic}
Rachaniotis, N.P., Dasaklis, T.K., Pappis, C.P.: A deterministic resource
  scheduling model in epidemic control: A case study. European Journal of
  Operational Research  \textbf{216}(1),  225--231 (2012)

\bibitem{ramirez2015point}
Ramirez-Nafarrate, A., Lyon, J.D., Fowler, J.W., Araz, O.M.:
  Point-of-dispensing location and capacity optimization via a decision support
  system. Production and Operations Management  \textbf{24}(8),  1311--1328
  (2015)

\bibitem{silhouette_score}
Rousseeuw, P.J.: Silhouettes: A graphical aid to the interpretation and
  validation of cluster analysis. Journal of computational and applied
  mathematics  \textbf{20},  53--65 (1987)

\bibitem{xiong2020impact}
Xiong, J., Lipsitz, O., Nasri, F., Lui, L.M., Gill, H., Phan, L., Chen-Li, D.,
  Iacobucci, M., Ho, R., Majeed, A., et~al.: Impact of {COVID}-19 pandemic on
  mental health in the general population: A systematic review. Journal of
  affective disorders  \textbf{277},  55--64 (2020)

\end{thebibliography}
\end{document}